# Single-Stage Regulated Resonant WPT Receiver with Low Input Harmonic Distortion

Kerui Li, *Student Member, IEEE*, Siew Chong Tan, *Senior Member, IEEE*, and Ron Shu Yuen Hui, *Fellow, IEEE*

*Abstract-* **Resonant rectifier topologies would be a promising candidate for achieving simple, compact, and reliable single-stage wireless power transfer (WPT) receiver if not for the lack of good DC regulation capability. This paper investigates the problems that prevent the feasibility of single-stage DC regulation in resonant rectifier topologies. A possible solution is the proposed differential resonant rectifier topology, of which the rectifier is designed to have a relatively constant AC voltage, and that phase-shift control is used to achieve relatively good output regulation. Design considerations on the reactive component sizing, magnetic component design, frequency and phase synchronization, small-signal modelling, and closed-loop feedback control design, are discussed. Experimental results verified that the proposed WPT receiver system can achieve single-stage AC rectification and DC regulation while attaining the key features of low harmonic distortion in its AC output voltage, continuous DC current, and zero-voltage-switching (ZVS) operation over a wide operating range.**

## I. INTRODUCTION

The wireless power transfer (WPT) technology is reshaping the battery charging technology of consumer electronics [1], [2]. It enables a more flexible, convenient, and safer method to charge portable electronic devices. Conventionally, research efforts are mainly focused on WPT transmitters [3]-[5]. Nevertheless, WPT receivers are also a critical part of the WPT charging system and deserves more research attention. The work presented in this paper is focused on the WPT receiver used in consumer electronics, of which the requirements of the WPT receiver system would be a simple circuit structure, compact size, and high reliability.

In typical WPT receiver systems, a combination of passive diode rectifier and post DC regulator is utilized [6]. Despite the fact that such a receiver system eases design complexity, the two-stage architecture requires a high component count, while the cascaded power processing introduces extra power loss and lowers the power conversion efficiency. In addition, the discontinuous conduction mode (DCM) operation of the diode rectifier at light load may lead to considerable current surge and overcurrent of the transmitter side switches, causing reliability issues.

Subsequently, the active full-bridge or half-bridge rectifiers are proposed [7]-[10], serving the purpose of elimination of post regulator in WPT receiver system and further simplification of the circuit configuration. However, they suffer from the disadvantages of having a high component count, complicated gate driver circuit, and low reliability. Furthermore, as the gate driving signal of the highside switch is connected to a floating switching node other than ground, auxiliary isolated gate driving circuits or bootstrap circuits are required. Moreover, the synchronization of the high-side and low-side switches of the bridge leg becomes very challenging at high switching frequency. Improper synchronization and present of switching noise may lead to a direct short circuit of the DC terminals, which damages the system. Adding dead time between the high-side and low-side PWM signals may prevent some of the faults from happening but is not a foolproof solution. In addition, the AC square waveforms produced by the bridge rectifier inherently contains rich harmonic components. This increases the total harmonic distortion present and necessitates the use of large EMI filters.

To address the aforementioned issues, the single-switch resonant topologies are utilized [11], [12]. In these solutions, the number of required semiconductor devices and reactive components are minimized. The use of ground-referenced power switch can prevent complicated gate driving circuity, direct short circuit risk and addresses the reliability issue. However, with these solutions, the zero-voltage-switching (ZVS) operation condition is highly sensitive to the operating line voltage and load. ZVS operation may be lost if operation condition changes. Moreover, the AC voltage waveform of the circuits contains both even and odd order harmonic components. This limits their THD performance and the size of EMI filter is substantially large.

The differential resonant topologies [13], [14] would be promising alternatives as they feature good THD performance and a low DC current ripple, which is beneficial to the size reduction of the required EMI filter. Nevertheless, they are overly sensitive to operate for variable load and coupling conditions [15]. Moreover, such solutions do not provide for single-stage DC regulation and a post regulator is required for the purpose.

In this paper, our investigation on development of a simple, compact and reliable WPT receiver, is reported. The proposed receiver system utilizes the differential resonant topology, which can provide low THD AC waveforms, continuous DC current, and ground-referenced switches. The low THD AC waveforms can reduce the size of the EMI filter, while the continuous DC current can reduce the required size of the output capacitor and also extend the lifespan of the associated battery. Phase-shift control [16] is adopted to realize the single-stage DC regulation in the active resonant WPT rectifier topology. To facilitate the DC regulation, the reactive components of the converter are properly designed such that the shape of the AC voltage waveforms maintains relatively constant for variable operating conditions. With the relatively constant AC waveforms, the received AC real power can be controlled by independently adjusting the phase-shift angle between the AC voltage and current waveform. By controlling the transmitted power, the DC output regulation is achieved.

## II. THE PROPOSED SINGLE-STAGE REGULATED RESONANT WPT RECEIVER SYSTEM

*A. Hardware Architecture*

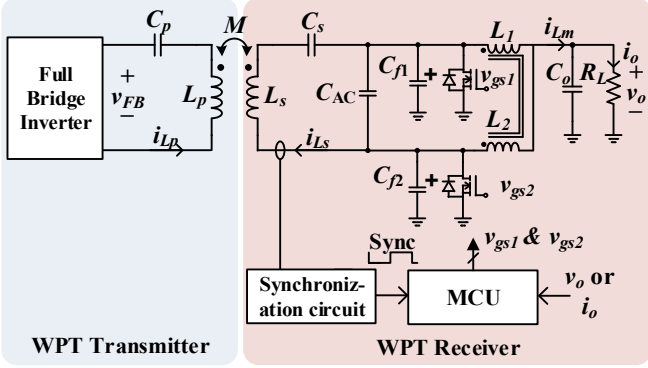

Fig. 1. Schematic diagram of the proposed WPT receiver system.

Fig. 1 shows the proposed WPT receiver system, whereby the conventional WPT transmitter topology [2] with series compensated WPT transmitter coil is utilized to provide high-frequency AC power to the WPT receiver. The receiver system comprises the series compensated WPT receiver coil ($L_s$ and $C_s$), an active differential class E rectifier power circuit with integrated coupled inductor, a synchronization circuit (a current sensor and an analog comparator) and a microprocessor (MCU). The inductance of the coil $L_s$ and the compensated capacitor $C_s$ are designed to resonate at switching frequency. The coupled inductors $L_1$ and $L_2$ have identical inductance, i.e., $L_1=L_2=L$, of which their mutual inductance $L_m$ equals to $kL$ ($0 \leq k \leq 1$). The capacitance of capacitors $C_{f1}$ and $C_{f2}$ are identical, i.e., $C_{f1}=C_{f2}=C_f$. The equivalent capacitance $C_{AC}+C_f$ is much larger than that of $C_s$ such that the resonance operation of the rectifier and the resonance of the series compensated wireless power coil are separated and will not affect one another. The AC voltage waveform of the system is designed to be relatively constant and nearly sinusoidal regardless of the varying coupling coefficient and load. This ensures the ZVS operation of the rectifier circuit over wide operation range, and is a critical factor for achieving single-stage AC rectification and DC regulation with phase-shift control. The differential architecture of the class E rectifier eliminates the even-order harmonics of the AC voltage waveform, while the odd-order current ripples of the coupled inductor are cancelled. The ground-referenced switches are operating complementarily.

*B. Operating Principles*

Analysis of the proposed WPT receiver involves the following assumptions:
1) The equivalent resistance of the reactive components are sufficiently small, and thereby negligible.

2) The output capacitor $C_o$ is sufficiently large such that its output voltage is relatively constant.

Due to the use of series-series compensation, the current of the WPT receiver coil $i_{Ls}$ is dependent on the transmitter AC voltage and the mutual inductance between the wireless power coils [17]. Consequently, the amplitude $|I_{Ls}|$ of the current $i_{Ls}$ is regarded as a constant value and $i_{Ls}$ is modelled as an independent sinusoidal current in the operational analysis of the WPT receiver system.

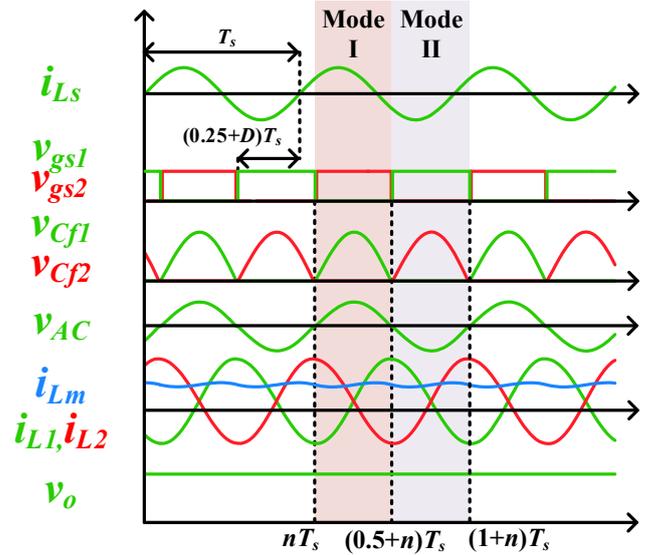

Fig. 2. Key waveforms of the WPT receiver system.

The key operation waveforms of the receiver WPT system is depicted in Fig. 2. The synchronization circuit detects the phase and frequency of $i_{Ls}$ and, thereby, generates a synchronization square pule *sync* to the MCU. After receiving the synchronization signal, a pair of complementary PWM ($v_{gs1}$ and $v_{gs2}$) with the switching frequency of $f_s$, the duty cycle of 50%, and the phase-shift ratio of $D$ ($0 \leq D \leq 0.25$), are produced by the MCU. The phase-shift between $v_{gs1}$ and $i_{Ls}$ is $(0.25+D)T_s$. Since the switches are operating complementarily, there are two operation modes for the power receiver circuit. The time interval of Mode I is between $nT_s$ to $(n+0.5)T_s$, while the time interval of Mode II is between $(n+0.5)T_s$ to $(n+0.5)T_s$, for $n$ being an integer number. The equivalent circuits of these two operation modes are shown in Fig. 3 for further elaboration and analysis.

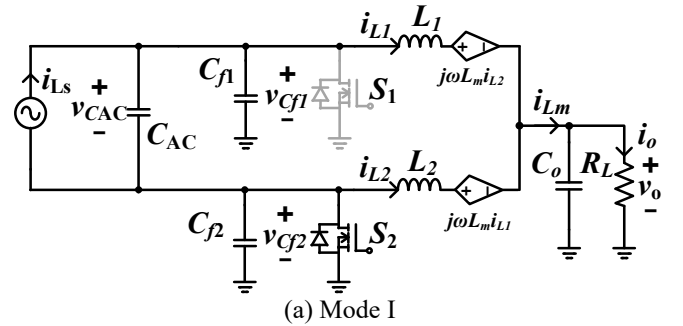

(a) Mode I

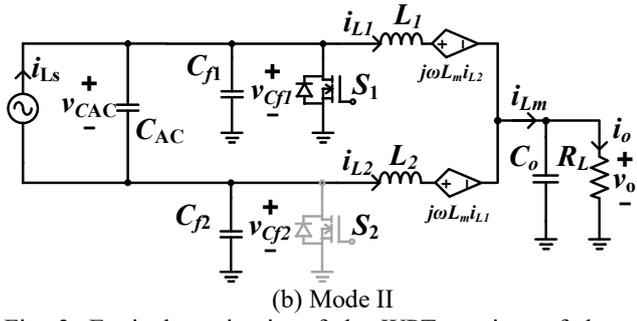

(b) Mode II

Fig. 3. Equivalent circuits of the WPT receiver of the two operation modes.

*Mode I ($nT_s \leq t < (n+0.5)T_s$)*

In Mode I (see Fig. 3(a)), switch $S_1$ is off, while switch $S_2$ is on. As a result, the voltage of capacitor $C_{f2}$ is zero, i.e.,

$$v_{Cf2} = 0 \qquad (1)$$

Inductor $L_2$ provides its stored energy to the output capacitor $C_o$ as

$$L_2 \frac{\partial i_{L2}}{\partial t} + kL \frac{\partial i_{L1}}{\partial t} = -v_o \qquad (2)$$

Meanwhile, capacitors $C_{f1}$ and $C_{AC}$ are being charged up by input current source $i_{Ls}$ and inductor $L_1$ as

$$(C_{AC} + C_{f1}) \frac{\partial v_{Cf1}}{\partial t} = i_{Ls} - i_{L1} \qquad (3)$$
$$= |I_{Ls}| \cos(2\pi f_s t - 2\pi D) - i_{L1}$$

Inductor $L_1$ resonates with capacitor $C_{f1}$ and $C_{AC}$ as

$$L_1 \frac{\partial i_{L1}}{\partial t} + kL \frac{\partial i_{L2}}{\partial t} = v_{Cf1} - v_o \qquad (4)$$

The resonant frequency of the tank formed by $C_{AC}$, $C_{f1}$, $L_1$, and $L_2$ is very close to the switching frequency $f_s(=1/T_s)$. The voltage waveform of $v_{Cf1}$ is sinusoidal. It rises up from zero volt at $t=nT_s$, and returns to zero volt at time $t=(n+0.5)T_s$ after a half cycle of resonance, providing the ZVS turn on and off conditions to switch $S_1$.

The output capacitor buffers the currents $i_{L1}$ and $i_{L2}$ from the coupled inductor and holds the output voltage as

$$C_o \frac{\partial v_o}{\partial t} = i_{L1} + i_{L2} - \frac{v_o}{R} \qquad (5)$$

*Mode II ($(n+0.5)T_s \leq t < (n+1)T_s$)*

In Mode II (see Fig. 3(b)), switch $S_2$ turns off and switch $S_1$ turns on at time $t=(n+0.5)T_s$. The voltage of capacitor $C_{f1}$ becomes zero, i.e.,

$$v_{Cf1} = 0 \qquad (6)$$

The drain terminal of switch $S_1$ is connected to the ground of the system. Inductor $L_1$ transfers its stored energy to output capacitor $C_o$ as

$$L_1 \frac{\partial i_{L1}}{\partial t} + kL \frac{\partial i_{L2}}{\partial t} = -v_o \qquad (7)$$

Concurrently, capacitors $C_{f2}$ and $C_{AC}$ are charged up by input current source $i_{Ls}$ and inductor $L_2$ as

$$(C_{AC} + C_{f2}) \frac{\partial v_{Cf2}}{\partial t} = i_{Ls} - i_{L2} \qquad (8)$$
$$= |I_{Ls}| \cos(2\pi f_s t - 2\pi D) - i_{L1}$$

Inductor $L_1$ resonates with capacitors $C_{f1}$ and $C_{AC}$ as

$$L_2 \frac{\partial i_{L2}}{\partial t} + kL \frac{\partial i_{L1}}{\partial t} = v_{Cf2} - v_o \qquad (9)$$

The resonant frequency of the resonant tank formed by $C_{AC}$, $C_{f2}$, and the coupled inductors is very close to the switching frequency $f_s(=1/T_s)$. The voltage waveform of $v_{Cf1}$ is sinusoidal. The voltage is increased from zero at $t=nT_s$, and returned to zero at time $t=(n+0.5)T_s$ due to the resonating operation, achieving both ZVS on and off operations for switch $S_2$.

The output capacitor absorbs the ripples of current $i_{Lm}$ and provides a fairly constant DC output current and output voltage, such that

$$C_o \frac{\partial v_o}{\partial t} = i_{Lm} - \frac{v_o}{R} = i_{L1} + i_{L2} - \frac{v_o}{R} \qquad (10)$$

By solving the aforementioned differential equations, the steady-state time-domain equations of the receiver are derived. The voltage waveforms of the capacitors $v_{AC}$, $v_{f1}$ and $v_{f2}$ are obtained as

$$v_{Cf1}(t) \approx \begin{cases} V_m \cos\left(\frac{(t-0.25T_s)}{\sqrt{L(C_{AC}+C_f)(1-k^2)}}\right) + (1-k)v_o & \text{if } nT_s \leq t \leq (n+0.5)T_s \\ 0 & \text{if } (n+0.5)T_s \leq t \leq (n+1)T_s \end{cases} \qquad (11)$$

$$v_{Cf1}(t) \approx \begin{cases} 0 & \text{if } nT_s \leq t \leq (n+0.5)T_s \\ V_m \cos\left(\frac{(t-0.75T_s)}{\sqrt{L(C_{AC}+C_f)(1-k^2)}}\right) + (1-k)v_o & \text{if } (n+0.5)T_s \leq t \leq (n+1)T_s \end{cases} \qquad (12)$$

$$v_{AC}(t) = v_{Cf1}(t) - v_{Cf2}(t) =$$
$$\begin{cases} V_m \cos\left(\frac{(t-0.25T_s)}{\sqrt{L(C_{AC}+C_f)(1-k^2)}}\right) + (1-k)v_o & \text{if } nT_s \leq t \leq (n+0.5)T_s \\ -V_m \cos\left(\frac{(t-0.75T_s)}{\sqrt{L(C_{AC}+C_f)(1-k^2)}}\right) + (1-k)v_o & \text{if } (n+0.5)T_s \leq t \leq (n+1)T_s \end{cases}$$
$$\approx 2\gamma v_o \sin(2\pi f_s t) \qquad (13)$$

where

$$V_m = \left|(1-k)\sec\left(\frac{0.25T_s}{\sqrt{L(C_{AC}+C_f)(1-k^2)}}\right)\right| v_o \qquad (14)$$

$$\gamma = \frac{2(1-k)}{\pi(1-4\pi^2 f_s^2 L(C_{AC}+C_f)(1-k^2))} \qquad (15)$$

To simplify the analysis, the value of $\gamma$ can be approximated as 1.58. The received AC real power of the receiver is then calculated as

$$P = \frac{1}{T_s}\int_{nT_s}^{(n+1)T_s} v_{AC} i_{Ls}\, dt = P_{max}\sin(2\pi D) \qquad (16)$$

where

$$P_{max} = \gamma|I_{Ls}|v_o \approx 1.58|I_{Ls}|v_o \qquad (17)$$

The receiver real power is transferred to the load resistor without loss. The output voltage and current are calculated as

$$P = v_o i_o = \frac{v_o^2}{R_L} \qquad (18)$$

By substituting (15) and (16) into (17), the output voltage and current of the receiver are calculated as

$$\begin{cases} i_o \approx 1.58|I_{Ls}|\sin(2\pi D) \\ v_o \approx 1.58|I_{Ls}|R_L\sin(2\pi D) \end{cases} \quad (19)$$

Eq. (19) shows that output current $i_o$ and output voltage $v_o$ can be regulated by adjusting $D$ independently.

As shown in Fig. 4, $i_o$ is dependent only on the phase-shift ratio $D$. For fixed phase-shift ratio, the output current is relatively constant, which is suitable for applications that require constant output current. However, for output voltage $v_o$ (see Fig. 5) it is highly load dependent. As a result, to obtain a relatively constant output voltage against load change, the phase-shift ratio $D$ has to be adjusted properly by using a feedback control loop.

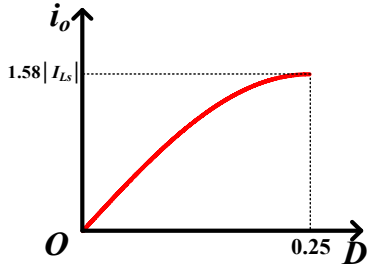

Fig. 4. Output current $i_o$ against phase-shift ratio $D$.

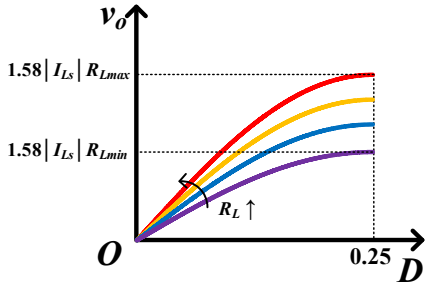

Fig. 5. Output voltage $v_o$ against phase-shift ratio $D$.

### III. DESIGN CONSIDERATIONS

*A. Design of the Rectifier Power Circuit*

The major design objectives of the reactive components in the rectifier are: 1) to attain relatively constant AC voltage waveform of $v_{AC}$ regardless of loads and phase-shift ratio; 2) to reduce the current ripples of the current $i_{LM}$; and 3) to enforce ZVS operation for variable loads and coupling coefficient.

The first design objective is to attain a relatively constant AC voltage waveform. Voltage $v_{AC}$ involves the contribution of voltage $v_{Cf1}$ and $v_{Cf2}$. To ensure a relatively constant voltage waveform of $v_{Cf1}$ and $v_{Cf2}$, the voltage components contributed by the coupled inductor $L_1$ and $L_2$ should be sufficiently larger than that by current $i_{Ls}$. However, this may lead to unnecessary reactive and conduction loss. Consequently, the effective resonant inductance of the coupled inductor is designed as

$$\frac{v_o}{40 f_s |I_{Ls}|} < L(1-k^2) < \frac{v_o}{4 f_s |I_{Ls}|} \quad (20)$$

The second design objective is to reduce the ripple of $i_{Lm}$. Assuming the ripple should be less than $x\%$ of the nominal value output current $i_o$, the boundary of the coupled inductance is designed as

$$kL > \frac{0.105 v_o}{x\% i_o f_s} \quad (21)$$

Feasible values of $L$ and $k$ can be selected by solving equations (20) and (21). The shaded area in Fig. 6 is the feasible region that fulfils both the equations. Proper values of the coupled inductor (the coupling coefficient $k$ and inductance $L$) can therefore be selected from the figure.

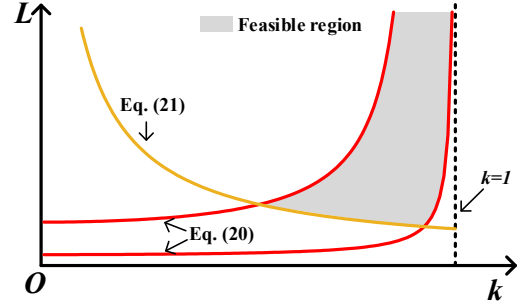

Fig. 6. Feasible region of the coupled inductor.

After obtaining the coupling coefficient $k$ and inductance $L$, the effective capacitance is designed. The third design objective is to ensure the ZVS operation. To attain ZVS operation, the voltage waveform of $v_{Cf}$ should be zero at time $0.5T_s$ and $T_s$, i.e., $v_{Cf1}(nT_s)=v_{Cf2}(nT_s)=v_{Cf1}((n+0.5)T_s)=v_{Cf2}((n+0.5)T_s)=0$. Meeting ZVS operation requirement (2nd design objective), the effective resonant capacitance is then designed as

$$(C_{AC} + C_f) = \frac{4\pi^2 f_s^2 \alpha^2}{L(1-k^2)} \quad (22)$$

where $\alpha$ ($1<\alpha<2$) is the root of the transcendental equation shown as follows

$$\frac{1-k}{1+k}\tan(\frac{\pi}{2}\alpha) + \frac{\pi}{2}\alpha = 0 \quad (23)$$

Selected numerical results for Eq. (23) are shown in Table I. The voltage stress of capacitor $C_{AC}$ is much higher than that of capacitors $C_{f1}$ and $C_{f2}$. Thus, it is recommended to use a capacitor $C_{AC}$ with lower capacitance.

TABLE I.
SELECTED NUMERICAL RESULTS

| k | α |
|---|---|
| 0 | 1.29 |
| 0.1 | 1.25 |
| 0.2 | 1.21 |
| 0.3 | 1.18 |
| 0.4 | 1.15 |
| 0.5 | 1.12 |
| 0.6 | 1.09 |
| 0.7 | 1.07 |
| 0.8 | 1.04 |
| 0.9 | 1.02 |

*B. Magnetic Component Design*

The coupled inductor has two design variables (coupling coefficient $k$ and inductance $L$) to be determined. The specific values are obtained using Eq. (20) and (21). However, the specific magnetic design cannot be reflected directly from the inductance and coupling coefficient. In this section, the magnetic component design is discussed.

Litz wire is used to wind the coupled inductor in order to overcome the skin effect. A magnetic core with three legs is used, as shown in Fig. 7. Both primary and secondary side windings have $n_1$ turns on the center leg, and $n_2$ turns on the outer leg. To facilitate the magnetic design, the magnetic circuit of the coupled inductor is shown in Fig. 8. $\mathcal{R}_1$ and $\mathcal{R}_2$ represent respectively the equivalent reluctance of centre legs and outer legs. The sources in the magnetic circuit are the magnetomotive forces introduced on the corresponding windings. The inductance $L$ and coupling coefficient $k$ are therefore obtained as

$$L = \frac{2\mathcal{R}_2 n_1^2 + 2\mathcal{R}_2 n_1 n_2 + (\mathcal{R}_1 + \mathcal{R}_2)n_2^2}{(2\mathcal{R}_1 + \mathcal{R}_2)\mathcal{R}_2} \quad (24)$$

$$k = 1 - \frac{n_2^2}{\mathcal{R}_2 L} \quad (25)$$

Eq. (24) and (25) show that to properly select turns $n_1$ and $n_2$, the desired values of inductance $L$ can be attained by adjusting the air gaps, while the coupling coefficient $k$ can be attained by adjusting the turns of windings.

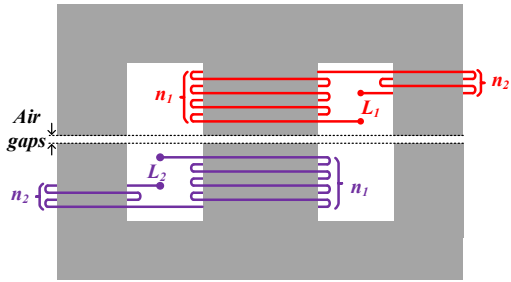

Fig. 7. Coupled inductor core and winding.

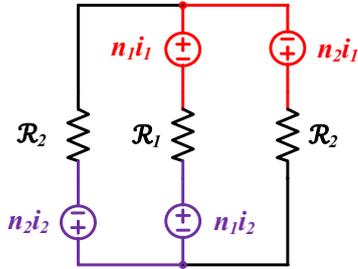

Fig. 8. Equivalent magnetic circuit.

### C. Frequency and Phase Synchronization

The implementation of synchronization and phase-shift control mainly comprises two major elements: frequency synchronization and phase-shift control. Fig. 9 shows the key waveforms and block diagrams of these two elements.

The top halves of Fig. 9(a) and Fig. 9(b) show respectively the waveform and block diagram of frequency synchronization. An external zero-crossing detector converts the sinusoidal signal of $i_{Ls}$ into a signal *Sync* that has square waveforms. *Sync* is then sent to the microcontroller TMS320F28335 via GPIO32, which is assigned for PWM synchronization [18]. The PWM1 module (see Fig. 9(b)) is programmed for frequency synchronization. The rising edge of *Sync* triggers the time base counter of PWM1 module TBCRT1 counting from zero cycle-by-cycle (see Fig. 9 (a)). The duty cycle of PWM1 is set at 50%. As a result, the frequency deviation between PWM1 and $i_{Ls}$ is minimized cycle-by-cycle. PWM1 is synchronized with $i_{Ls}$ with a fixed phase-shift of $0.25T_s$.

The bottom halves of Fig. 9(a) and Fig. 9(b) show respectively the waveform and block diagram of the implementation of phase-shift control. Taking the synchronized PWM1 as reference (by using the internal synchronization signal *InSync* from PWM1), PWM 2 module applies the phase-shift ratio $D$ to its time base counter TBCTR2, where the phase-shift ratio $D$ is obtained from the compensator. Hence the carrier waveform TBCTR2 has a phase-shift of $DT_s$ against the PWM1 time base counter TBCTR1. The duty cycle of PWM2 is set at 50%. By utilizing TBCTR2 and the duty cycle, the required complementary phase-shift PWM signal $v_{gs1}$ (CMP2>TBCTR2) and $v_{gs2}$ (CMP2<TBCTR2) are generated. The phase difference between PWM1 and $v_{gs1}$ is $DT_s$, and thus the total phase shift angle between $v_{gs1}$ and $i_{Ls}$ is $(0.25+D)T_s$ (see Fig. 9(a)). The generated phase-shift PWM $v_{gs1}$ and $v_{gs2}$ are then used to drive the circuit into regulating the DC output voltage.

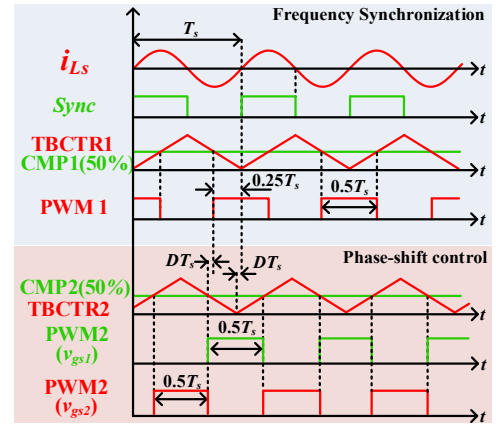

(a) Key waveforms

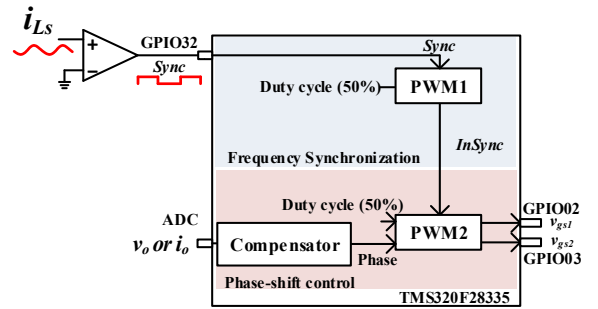

(b) Block diagrams

Fig. 9. Key waveforms and Block diagram of frequency synchronization and phase-shift control.

### D. Small-Signal Model and Feedback Control

The output capacitor $C_o$ is operating as an energy buffer. The transmitted real power $P(D)$ is delivered to the load with buffering of the capacitor $C_o$ described as

$$\frac{\partial \frac{1}{2} C_o v_o^2}{\partial t} = P(D) - v_o i_o \quad (26)$$

To manipulate Eq. (29), the small-signal models of the output voltage and current can be obtained. The corresponding feedback control loops are designed accordingly.

*(i) Dynamics of Output Current*

The control variable is the output current $i_o$. By substituting Eq. (16) into (26) and manipulating the variable, the differential equation of $i_o$ can be written as

$$R_L C_O \frac{\partial i_o}{\partial t} = 1.58|I_{Ls}| \sin(2\pi D) - i_o \quad (27)$$

By linearizing Eq. (27), the small-signal dynamic model can be expressed as

$$R_L C_O \frac{\partial \widetilde{i_o}}{\partial t} = 9.93|I_{Ls}| \cos(2\pi D) \widetilde{D} - \widetilde{i_o} \quad (28)$$

The Bode plots of the open-loop transfer function (28), and the numerical results obtained from circuit simulation are shown in Fig. 10. The theoretical and the simulation results are very close at the low-frequency range of ≤1000 Hz. This validates the accuracy of the derived small-signal equation.

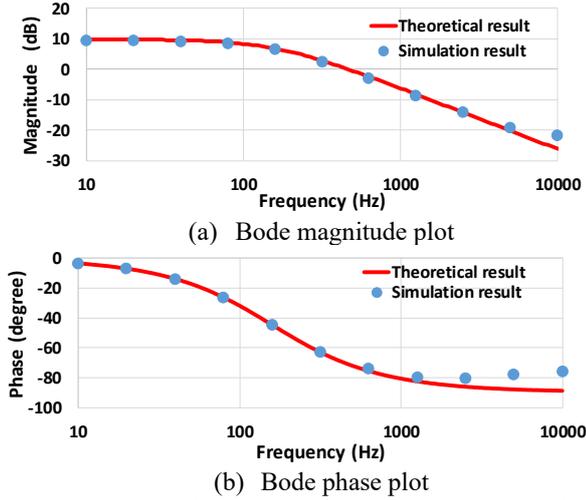

(a) Bode magnitude plot

(b) Bode phase plot

Fig. 10. Bode plots of the output current to the phase-shift ratio transfer function.

*(ii) Dynamics of Output Voltage*

Similarly, the differential equation of the output voltage can be derived as

$$R_L C_O \frac{\partial v_o}{\partial t} = 1.58|I_{Ls}|R_L \sin(2\pi D) - v_o \quad (29)$$

By linearizing (29) with consideration to the small-signal AC perturbation of $D$, the small-signal model of the output voltage is obtained as

$$R_L C_O \frac{\partial \widetilde{v_o}}{\partial t} = 9.93|I_{Ls}|R_L \cos(2\pi D) \widetilde{D} - \widetilde{v_o} \quad (30)$$

The Bode plots of the transfer function and the numerical simulation results are depicted in Fig. 11. The simulation results are very close to the theoretical (calculated) ones, except for those points at the high-frequency range.

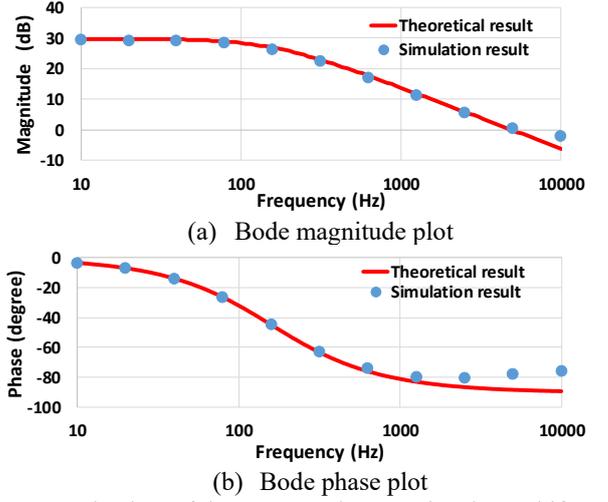

(a) Bode magnitude plot

(b) Bode phase plot

Fig. 11. Bode plots of the output voltage to the phase-shift ratio transfer function.

*(iii) DC Output Regulation*

To achieve good output regulation, the PI compensator, as shown in Fig. 12, is used. To avoid integral overflow, an anti-windup loop is added to the PI compensator.

For output current regulation, the corresponding values of $k_{pc}$ and $k_{ic}$ of the PI compensator are designed as

$$k_{pc} = \frac{f_{cc} C_O R_{Lnominal}}{1.58|I_{Ls}| \cos(2\pi D)}, k_{ic} = \frac{k_{pc}}{R_{Lnominal} C_O} \quad (31)$$

where $f_{cc}$ is the desired crossover frequency, $R_{Lnominal}$ is the nominal equivalent resistance of the nominal load, and $D$ is the nominal phase-shift ratio.

For output voltage regulation, the corresponding values of $k_{pv}$ and $k_{iv}$ of the PI compensator are designed as

$$k_{pv} = \frac{f_{cv} C_O}{1.58|I_{Ls}| \cos(2\pi D)}, k_{iv} = \frac{k_{pv}}{R_{Lnominal} C_O} \quad (32)$$

The control algorithm is implemented using a digital controller. Thus, $S$ domain parameters of the PI compensator are converted to $Z$ domain parameters using the Tustin $Z$ transform before implementation.

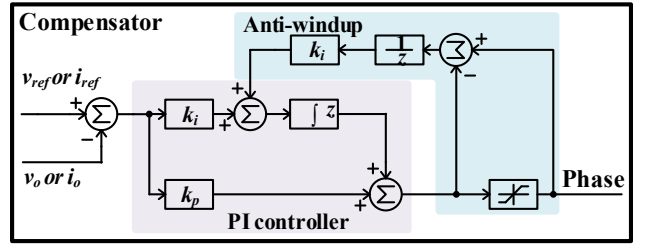

Fig. 12. Block diagram of the PI compensator.

## IV. EXPERIMENTAL RESULTS

A 200 kHz switching frequency WPT receiver prototype is built. The nominal output voltage is 12 V, the maximum output current is 2 A and the power is 24 W. The parameters and the components of the prototype are shown in Table II. A picture of the experimental setup and the prototype is shown in Fig. 13.

TABLE II.
LIST OF COMPONENTS

| Part | Value | Part Number |
|---|---|---|
| $C_{f1}, C_{f2}$ | 47 nF | |
| $C_{AC}$ | 2 nF | |
| $L_1, L_2$ | 22.8 µH, 21.7 µH | |
| $k$ | 0.71 | |
| Magnetic core | | B65815-E-R49 |
| $L_s$ | 164 µH ($d$=29 cm) | |
| $C_s$ | 3.86 nF | |
| $C_o$ | 6800 µF | |
| Gate Driver | | ADuM3223 |
| Comparator | | LM 393P |
| Current Transformer | | AS-100 |
| MOSFET | | IRFP 250M |
| Microcontroller | | TMS320F28335 |

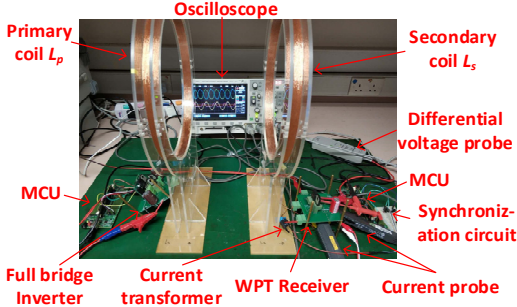

Fig. 13. Experimental setup and the prototype.

*A. Steady-State Performance*

The key waveforms of the receiver in steady state are shown in Fig. 14. The output voltage is 12 V and the current is 2 A. The AC voltage waveform $v_{AC}$ and current waveform $i_{Ls}$ are nearly in phase with one another ($D$=0.25).

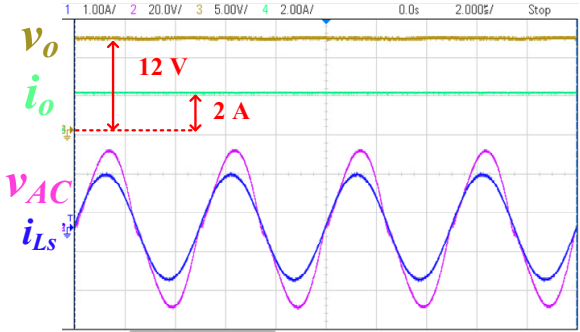

Fig. 14. Steady-state waveforms of the receiver.

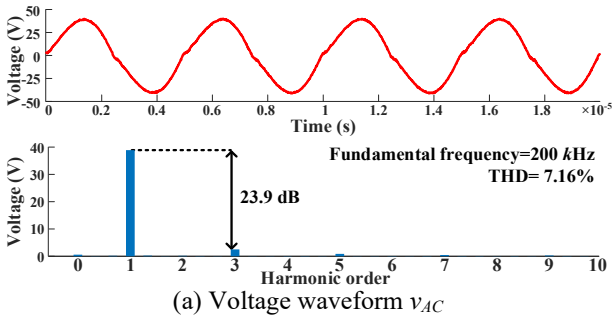

(a) Voltage waveform $v_{AC}$

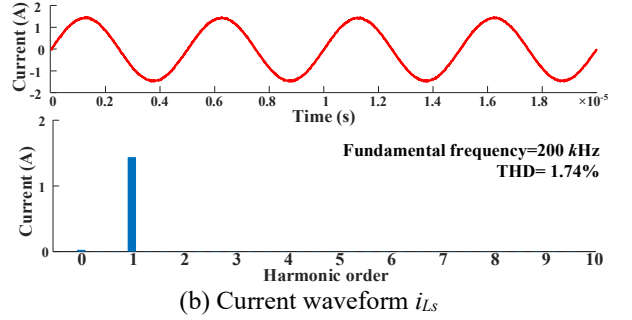

(b) Current waveform $i_{Ls}$

Fig. 15. FFT and THD analysis of the AC voltage waveform and current waveform.

Fig. 15 shows the Fast Fourier Transformation (FFT) and THD analysis of the waveform of $v_{AC}$ and waveform of $i_{Ls}$. The THD of $v_{AC}$ is 7.16%. The even-order harmonic components are eliminated. The third harmonic component, which has the highest amplitude among all harmonic components, is attenuated by 23.9 dB as compared to the fundamental component. Owing to the much-reduced distortion of the voltage waveform, the waveform of resonant current $i_{Ls}$ possesses a THD of only 1.74%. The high-order harmonic components in the spectrum are almost eliminated.

Fig. 16 shows the ZVS operation of the switches of the receiver. As shown in Fig. 16(a), the voltage waveform of switch $S_1$ increases after the gate driving signal turns off, resulting in ZVS turn off. On the other hand, the voltage waveform drops to zero before $v_{gs}$ turns on, leading to ZVS turn on. Similarly, the waveforms in Fig. 16 (b) validate the ZVS turn on and turn off operations of switch $S_2$. Thus, there is virtually no switching loss on the rectifier.

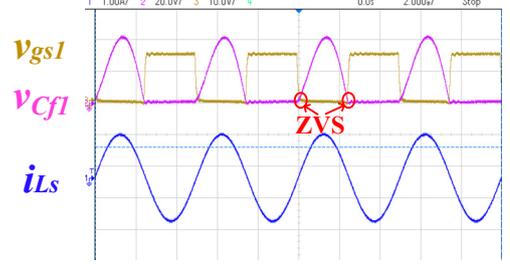

(a) ZVS operation of $S_1$

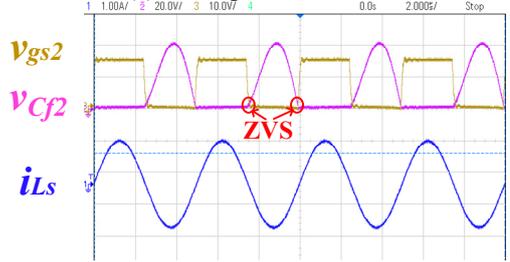

(b) ZVS operation of $S_2$

Fig. 16. ZVS turn on and off operation of the switches.

Fig. 17 shows the currents of the coupled inductor. The peak-to-peak value of the coupled inductor currents under operation is around 6.4 A while that of the resonant current $i_{Ls}$ is around 0.8 A. The reactive circulating current facilitates the relatively constant waveform of $v_{AC}$, $v_{Cf1}$ and $v_{Cf2}$. The differential architecture significantly reduces the peak-to-peak current ripples to 0.8 A, which is around 40% of the maximum output current. The continuous current flow of $i_{LM}$ with small ripple reduces the required capacitance needed on the output capacitor.

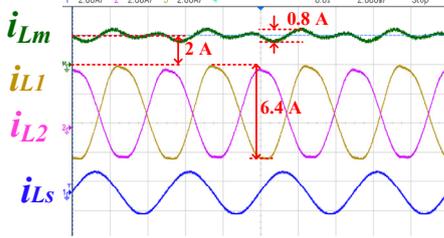

Fig. 17. Current of the coupled inductor.

Fig. 18 shows the measured efficiency of the receiver system. Without on-off keying control, a maximum efficiency of 90.2% is achieved at full load. The efficiency drops as the load power decreased. At light-load operation, the phase-shift ratio is relatively small. Consequently, the AC voltage and current waveforms are out of phase, which causes significant reactive power to be generated, thereby lowering efficiency.

To increase the efficiency at light load, the on-off keying control on the transmitter is used [19]. With on-off keying control, the full-load efficiency remains at 90% while the light-load efficiency increases by around 15%. The on-off keying control reduces the amplitude of the AC current, which leads to a relatively large phase-shift ratio. Hence, the reactive power is reduced, lowering the generated conduction loss and resulting in higher efficiency at light load operation.

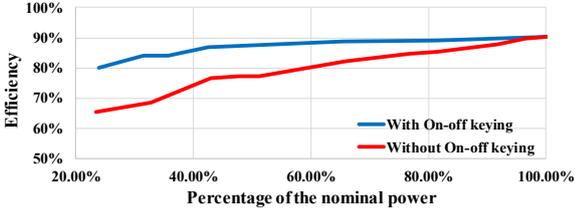

Fig. 18. Efficiency of the receiver system.

The measured output voltage of the rectifier for different load power and coil's distance under output voltage regulation is shown in Fig. 19. The reference voltage is 12 V. The maximum steady-state voltage regulation error is 0.1 V, which is around 0.84% of the reference voltage as the power varies from 20% load power to full power. The maximum voltage regulation error is 0.1 V (0.84% of the reference voltage) as the distance between the coil increases from 9 cm to 17 cm.

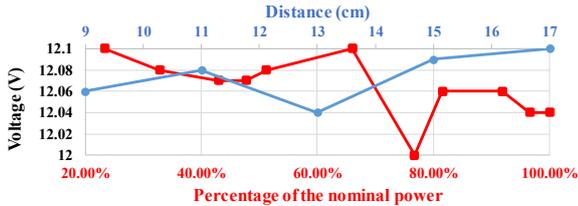

Fig. 19. Output voltage of rectifier versus nominal power and coils' distance under output voltage regulation.

Fig. 20 shows respectively the measurement results of the output current of the rectifier against the load resistance and the coils' distance under output current regulation. The reference current is 1 A. The maximum steady-state current regulation error is 0.04 A (0.4% of the reference current). The maximum current regulation error is 0.06 A (0.6% of the reference current) as the distance varies from 9 cm to 17 cm. The results validate that the WPT rectifier system is well-regulated by the proposed feedback voltage and current control schemes.

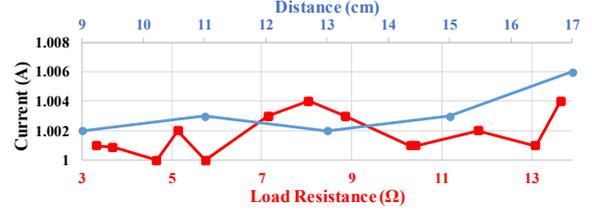

Fig. 20. Output current of rectifier versus load resistance and coils' distance under output current regulation.

*B. Dynamic Performance*

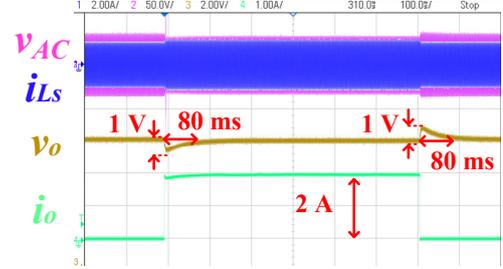

Fig. 21. Waveforms of rectifier operating with step-current changes under output voltage regulation.

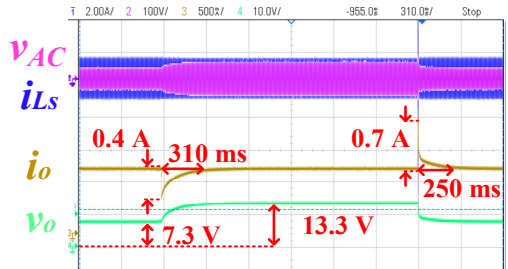

Fig. 22. Waveforms of rectifier operating with step-resistance changes under output current regulation.

Fig. 21 depicts the waveforms of the rectifier operating with respect to step-current changes under output voltage regulation. With the current step changing from 0 A to 2 A, the output voltage has a 1 V dip (8.3% of the reference output voltage). With the current step changing from 2 A to 0 A, the resulting overshoot is also 1 V. The output voltage took approximately 80 ms to return to steady-state value after each step-current change.

Fig. 22 depicts the waveforms of the rectifier operating with respect to step-resistance changes under output current regulation. The output current has a 0.4 A dip after the resistance steps from 13.3 Ω to 7.3 Ω. The settling time is around 310 ms. For resistance stepping from 7.3 Ω to 13.3 Ω, the output current has an overshoot of 0.7 A. The settling time is around 250 ms. The dynamic performance of the system with respect to the load-step changes are satisfactory for general battery charging operations.

*C. Comparative Study*

TABLE I.
A COMPARISON WITH EXISTING WPT RECEIVERS

|  | **Proposed** | Qi compliant receiver [6] | [8] | [12] |
|---|---|---|---|---|
| Topology | **Active Differential Class E rectifier** | Diode rectifier & Linear regulator | Active Class D rectifier | Active Class E rectifier |
| Power Processing Stages | **Single stage** | **Two stage** | **Single stage** | **Single stage** |
| Compensation of Receiver Coil | **Series compensation** | Series parallel compensation | Series compensation | Series compensation |
| Resonant Frequency | **200 kHz** | 100 kHz | 917 kHz | 200 kHz |
| Regulation Frequency | **200 kHz** | N/A | 91.7 kHz to 917 kHz | 200 kHz |
| Regulation Objectives | **Voltage & Current** | Voltage & Current | Voltage | Voltage |
| Modulation | **Phase-shift Modulation** | Pulse Width Modulation | Pulse Density Modulation | Phase-shift Modulation |
| Number of Power MOSFETs | **2** | 1 | 2 | 1 |
| Number of Power Diode | **0** | 4 | 0 | 0 |
| Maximum Voltage stress | **≈3.14$V_o$** | ≈$V_{DC-link}$ | $V_o$ | ≈3.5$V_o$ |
| Theoretical THD | **≈4%** | ≈48% | ≈48% | ≈44% |
| Soft Switching Operation | **ZVS ON and OFF** | No | ZVS ON | ZVS ON and OFF |
| Output Current | **Continuous** | Continuous | Discontinuous | Continuous yet fluctuation |
| Output Voltage & Current | **12 V, 2 A** | 5.3 V, 1 A | 33 V, 1.32 A | 24 V, 0.47 A |
| Maximum Efficiency | 90% | 76% | N/A | 93% |

A comparison of the proposed system with existing WPT receivers on different aspects is provided in Table I. As illustrated, the two-stage solution [6] is less efficient as compared to the single-stage solutions, including the proposed one, [8], and [12]. The elimination of the post regulator further simplifies the architecture of the receiver system and has the potential to enhance efficiency. The proposed solution, as a single-stage solution, can achieve the lowest THD and largest output current capability amid the solutions. As compared with active class D rectifier solution [8], the proposed solution features continuous output current, constant regulation frequency and low THD, which can ease the design of filters and reduce output capacitance. As compared with active class E rectifier solution [12], the proposed solution characterizes slightly lower voltage stress and THD, which can ease of design. Additionally, the proposed solution presents highest output current capability, which shows the potential for future medium power WPT systems.

V. CONCLUSIONS

In this paper, a single-stage regulated resonant WPT receiver system with low input harmonic distortion, is presented. The receiver is based on the differential resonant rectifier topology and adopts the phase-shift control. An experimental prototype of the system has been constructed and tested for validation. Experimental results verify that low input harmonic distortion and accurate out voltage/current regulation of the receiver is achievable. The results also verify that satisfactory dynamic responses of the output voltage (under output voltage regulation) and output current (under output current regulation) can be achieved. The proposed receiver circuit seems promising as a potential solution for simple, compact and reliable WPT battery charging.

V. ACKNOWLEDGMENTS

This work was supported by the Hong Kong Research Grant Council under the GRF project 17204318.